# The Transformative Effects of AI on International Economics

Rafael Andersson Lipcsey[†]

April 23, 2023

**Abstract**

As AI adoption accelerates, research on its economic impacts becomes a salient source to consider for stakeholders of AI policy. Such research is however still in its infancy, and one in need of review. This paper aims to accomplish just that and is structured around two main themes; Firstly, the path towards transformative AI, and secondly the wealth created by it. It is found that sectors most embedded into global value chains will drive economic impacts, hence special attention is paid to the international trade perspective. When it comes to the path towards transformative AI, research is heterogenous in its predictions, with some predicting rapid, unhindered adoption, and others taking a more conservative view based on potential bottlenecks and comparisons to past disruptive technologies. As for wealth creation, while some agreement is to be found in AI's growth boosting abilities, predictions on timelines are lacking. Consensus exists however around the dispersion of AI induced wealth, which is heavily biased towards developed countries due to phenomena such as onshoring and reduced bargaining power of developing countries. Finally, a shortcoming of economic growth models in failing to consider AI risk is discovered. Based on the review, a calculated, and slower adoption rate of AI technologies is recommended.

---

[†] Email: rafael.anderssonlipcsey@sciencespo.fr

# 1. Introduction

It is clear, that artificial intelligence (AI) is a transformative force of a magnitude that will greatly affect the lives of humans in the decades to come. While media attention around the topic has truly ramped up only recently, AI technologies are already being employed within healthcare, retail, banking, manufacture and several other sectors. Having established the importance of AI in the years up ahead raises essential follow up questions in most fields of research, including economics, to which this paper belongs. What exactly is coming, and what does it imply for the worlds' economies and the field of economic theory? Is AI a force of infinite wealth to be welcomed, or is it just hype, to be treated with caution? This literature review formulates two research questions:

Firstly, it asks, how does the path towards transformative AI look like, and what is the timeframe within which we can expect to see such effects take place? Secondly, it inquires, how much wealth can AI create, and how will that wealth be shared?

Throughout the review, a specific focus will be dedicated to international trade for two reasons. It has been identified that the sectors most impacted by AI are also the most tradeable sectors. At the same time, research on AI's impacts through international value chains, and attempts at integrating AI into international trade theory is rather limited. As such, an international trade lens was deemed valuable.

# 2. How does the path towards transformative AI look like and what is the timeframe within which we expect such changes to take place?

**2.1 Integrating AI impacts into economic theory**

A first step in answering the questions above is briefly examining attempts at integrating AI impacts into economic theory in general, and trade theory in specific. A function popularly used in literature is the Cobb Douglas production function, which describes output as a function of labour and capital. Building on these theoretical fundamentals, Allan Dafoe (2018) created an AI production function, where AI development is deduced as a function of data, talent, compute, time, investment and other indicators such as prior progress and achievements. The use of a modified Cobb Douglas function of that nature however is questioned by some, due to its limits in accounting for shared dependencies and interactions between the various factors of production, or in this case, AI development. Langenkamp (2022) suggests the use of Wardley Maps instead which are built around a three-step process of describing the case, defining technological capabilities and finally ordering capabilities on a map. Perhaps the most ambitious attempt to date however comes from Trammell and Korinek (2023), who focus on three types of transformative AI impacts in economics, namely output growth, wage growth and on the labour share, aiming to bridge the gap between shorter term economic theory and more longtermist thinking using a variety of theoretical scenarios.

As for attempts at integrating AI impact into international trade theory, literature is still very limited. The only paper identified in this review is by Hazari et al. (2022) who use a Jones and Manuelli production function to introduce effects of automation and AI into an international trade framework. Both Trammell and Korinek (2023), as well as Hazari et al. (2022)'s work will be covered in more detail at a later stage in this review.

**2.2 Supply chain and sectoral perspectives from international trade**

When trying to predict the path of transformative impacts of AI on international trade, a supply chain/sectoral perspective is worth discussing. A twofold yet intertwined dilemma is presented: Firstly, what are the sectors which produce things that can be used as inputs to an AI based transformation? Secondly, which sectors are most likely to be enhanced by AI technology first, or put differently, in which sectors are AI innovations the most common? It is clear for example that the manufacturing sector will be heavily affected, as it is uses for instance sensors, IoT devices, advanced analytics and, now also AI, to function in the most efficient way possible (Manyika et al., 2017; Stackpole, 2020). At the same time, many of the outputs of the manufacturing sector, such as sensors are being increasingly used to aid the ever-increasing hunger of AI for data.

Process industries like agriculture, chemicals, metals and mining too, have a lot to benefit from AI technologies, such as AI powered demand sensing (Ferencz et al., 2022). These sectors don't necessarily produce direct inputs into AI, but their output moves through value chains to the manufacturing sectors, which then in turn supply AI with inputs. An enhancement of efficiency can also be seen in service supply chains, where AI offers personalized recommendations, customer support and other innovations (Ferencz et al., 2022).

A further important perspective concerns the observation that AI innovations are highest for sectors which are also the most tradable, hence the most deeply integrated into global value chains. As found by Ferencz et al. (2022), sectors with the highest foreign value added in final demand, a measure for global value chain integration, also have the highest patents, trademarks and publications share in AI innovations.

What can be concluded from the above? Firstly, we can expect AI impact to happen both through forward and backward linkages in global value chains. Furthermore, we have some idea of the sectors that will be impacted and will impact the AI transformation the most. We also know that AI and international trade are heavily interlinked, as the most tradeable sectors are the ones most impacted by AI. One key question however remains: How quickly can we expect the transformative impacts of AI to appear?

**2.3 Turning to past disruptions**

If we attempt to take a global value chain perspective in tackling the above question, using similar datasets from the OECD which Ferencz et al. (2022) used for tradability, we don't get very far. Databases such as the ICIO (Inter-Country Input Output Tables), and the Leontief input-output methodologies behind it can show the ultimate effects of shocks to the global economy (and individual countries and sectors) however they do not provide any time perspective of when these are expected to happen.

One option is to turn to past technological disruptions and assume that AI impacts will look somewhat similar. Mateos-Garcia (2020) theorizes that AI impacts will take longer to materialize than newspaper headlines make us believe. Jobs will not disappear completely, but will rather change and evolve over time, similarly to the evolution of the internet. He argues that prediction machines like AI do not only increase the amount of decisions to be made based on AI recommendations, but also the number of decisions which need to be made in order to integrate AI into our economies and societies in a safe and regulated manner. More about the decision-making bottleneck will be discussed below, however the comparison of the AI revolution to the impacts of the internet should be nuanced. On August 6, 1991 the World Wide Web as we know it became commercially available to consumers (Bryant (2011)), and it reached a hundred million users in a little over 8 years, or about 100 months later, on November 12, 1999 (The New York Times, 1999). Meanwhile, Bing AI, powered by Open AI's ChatGPT engine reached a 100 million users in about a month (Mashable, 2023). While the above is undeniably not a completely fair comparison, it may give us an indication regarding the sheer magnitude of the rate of AI adoption.

**2.4 Feedback loops**

Another way to attempt to quantify the leap towards transformative AI is by investigating feedback loops. One such loop is linked to the automation of computer production. If AI could theoretically replace all the labour needed in the production of computers, it could technically produce itself, which would fuel a loop leading potentially to exponential growth in adoption. This review did not find any research exploring such a scenario with a time frame included. According to a study by McKinsey Global Institute, half of today's work activities could be automated by 2055 (Manyika et al., 2017). In addition, it is predicted that physical activities, such as those found in manufacturing have the highest (81 percent) automation potential. Krishanswamy meanwhile explores the possibility of fully automating semiconductor assembly and testing. He concludes that full (level 5) automation is no longer an unreachable myth but provides no timeline for when this could be achieved. A last perspective stemming from the field of molecular biology comes from Kriegman et al (2021), who managed to create an AI designed, self-replicating and configurable organism called a xenobot, which marks one of the very first steps towards a future where self-replication of organisms is no longer a just a concept of science fiction.

Feedback loops however need to be limited to manufacturing but can also be knowledge based. Davidson (2021) explores how AI could affect growth over the long term. He introduces an additional feedback loop apart from the well-known *more ideas → more people → more ideas*, which has traditionally driven economic growth. The new loop, *more ideas → more AI systems → more ideas* could potentially lead to exponential growth, and thus significant impacts of AI on the world. Davidson (2021) assigns a higher than 30% probability that human-level AI has been developed in time for economic growth to reach 30 percent by the year of 2100. Furthermore, he theorizes that over a third of that growth would happen due to the birth of human-level AI.

**2.5 Bottlenecks**

A further approach that can be taken is looking at possible bottlenecks for rapid AI adoption. These, along with an evaluation of their solvability can provide us with valuable information regarding the feasibility of transformative AI within the next decades.

Bottlenecks can be broadly classed into 4 categories: supply chain, regulatory, technological, and knowledge based. To keep up with exponential economic growth, inputs to production such as raw materials (Davidson (2021), will need to be extracted, as well as intermediate products, such as chips and semiconductors will need to be produced, at a rate that can keep up. In addition, logistics lines will also need to be scaled up to a very significant extent. Next, as covered earlier, the number of decisions on how to safely integrate AI into our economies and societies will vastly increase (Mateos-Garcia, 2019). This also relates to Davidson (2021)'s argument regarding the possibility that humanity will simply decide to harness the potential of AI in a sustainable and more controlled way, even if it was possible to choose exponential growth. Furthermore, technical and technological limits, such as a roof to the type and percentage of tasks automatable by AI, or the requirement of technologies for upscaling that we do not yet possess could slow down progress (Davidson, 2021). Finally, perhaps too few new innovations will be made and too slowly, and perhaps we will lack the resources to upskill and re-skill people at a rate fast enough to keep up. All of the above could prove to be detrimental for fast paced, AI supported growth, however at the same time, many could be alleviated by AI technologies themselves.

**2.6 AI problems require AI solutions?**

When it comes to specific benefits of AI for economies, the literature is fairly vast, and some of the improvements discussed could alleviate, or completely eliminate bottlenecks discussed above. For instance, AI helps the development of more proactive and efficient global supply chains through its predictive capabilities (Jayathilaka, 2022; Achar, 2019; Ferencz et al, 2022), which may alleviate supply chain and logistical bottlenecks. AI is also used for boosting the efficiency of compliance software, streamlines contract creation (through simplifying legal language) (Jayathilaka, 2022; Jones, 2023), and bolsters trade negotiations (through advanced translation) (Achar), all of which could reduce regulatory and decision burdens. Furthermore, technological capabilities already are, and will be further improved using AI technology, such as significant advances in computing power and decreasing costs (Anderson & Rainie, 2018). Finally, AI is to an increasing degree used for education management and delivery. AI technologies can both be used to reduce regulatory and administrative burden on educators, freeing up precious time for teaching, as well as enhance student's learning through adaptive learning techniques (Miao et al., 2021).

# 3. How much wealth can transformative AI create and how will that wealth be shared?

Based on the first part of this review, it is clear, that while faced with large uncertainty, and several bottlenecks, AI has the potential to strongly boost the efficiency of global value and production chains and bring about strong economic growth. Is it possible however to predict how much wealth transformative AI can create and how that wealth will be shared?

**3.1 Scenarios of growth**

Predicting how much growth a technology in its beginning stages can bring to the world economy is near impossible, hence we must rely on scenario building and assumptions. Furthermore, outcomes vary widely depending on the type of model used. Both Davidson (2021) and Trammell and Korinek (2023) employ various models in their analyses, such as complete or partial substitution of AI into a production function, inclusion of AI into task-based functions, as well as standard and exogenous growth models. While a variety of scenarios are used, broadly speaking we can discern between three main ones: standard growth, explosive growth and stagnation. Standard growth is exponential, as has been over the last 150 years, meanwhile explosive growth can be much higher than that. Combined with historical evidence concerning the increasing pace of growth, and the above detailed idea-based feedback loops leads Davidson (2021) to assign a 10 percent probability to explosive growth occurring by 2100. Trammell and Korinek (2023) conclude that AI aided explosive, even super-exponential growth is possible for a certain time, until production reaches limiting factors such as energy production or land usage. Still, Trammell and Trammell and Korinek (2023) note that as of now, we can't put a timeframe to transformative impacts from AI, if these even were to take place.

Apart from standard and explosive growth scenarios, both Davidson (2021) and Trammell and Korinek (2023) consider possibilities of stagnation. Trammell and Korinek (2023) also introduce an interesting and relevant gap between futurist research and economic theory, namely, the risk of an AI induced existential catastrophe. It is clear, that economic literature lacks discussion on this topic, as argued, not because of a lack of interest, but because explosive improvements in automation and capital productivity plugged into existing growth models are exclusively beneficial, and don't pose a risk. They theorize, that the key here is the degree to which AI will replace human labour. As long as human labour will remain a bottleneck for AI growth, and AI self-replication feedback loops won't take place, humans will maintain control over AI, and not vice versa.

Finally, the most concrete attempt at predictions of growth comes in the form of a 2018 discussion paper by Bughin et al. from McKinsey Global Institute. By 2030, the value-added gains boost stemming from AI is predicted to be 26 percent, with externalities and transition costs bringing the net impact to 16 percent. An important point brought forward in the paper is the magnifying power of competitive pressure on AI adoption. By 2030, competitive pressure is expected to boost AI adoption by 13 percentage points, from 25 to 48 percent (Bughin et al., 2018).

**3.2 Trade frictions: lessons from the present and past**

Many of the benefits that AI brings to global value chains have already been discussed above. Most of the examples brought up all have one common theme: reduction of trade frictions. Thus, taking a trade frictions perspective in attempting to answer how much wealth transformative AI can bring to the world economy is worthwhile.

In looking at the scene of online trade, some conclusions can be drawn. For example, it has been found that trade frictions caused by physical geographic distance when it comes to trade in goods are lower for online trade than for offline trade, with distance coefficients nearly halved (for the EU) of -0.747 for online, and -1.294 for offline trade (Gomez-Herrera et al, 2014). A study on a Norwegian dataset meanwhile found that a high degree of internet availability could actually lead to a slight increase in elasticity of trade when it comes to distance, and thus make trade even more sensitive to distance than it was before (Akerman et al., 2018). It seems, given the above, that even when it comes to more historic technological advances such as the adoption of internet, consensus is lacking on whether, and if so, how much such technologies have reduced trade frictions. Much of this can most likely be attributed to the sheer magnitude and diversity of the online trading space, and the lack of datasets that might be used to tackle this problem.

That being said, even for AI, specific use cases can provide proof of concept for its potential in reducing trade frictions, specifically when it comes to AI language translation models. Brynjolfsson et al, (2018) found that introduction of improved machine translation systems could increase international trade on eBay by 10.9 percent, a clear testament for the usefulness of AI in this specific case. While research in this field is still in its infancy, it can be expected that an increasing number of studies relating to specific cases will be published in the years to come. Provision of a macro perspective will however surely prove to be a difficult challenge.

### 3.3 Sectoral divide

When trying to answer the question of how AI induced wealth will be shared, it is a good exercise to take a sectoral perspective first. It is clear, as one might also have read in one of numerous media articles, that some sectors will benefit more from AI innovations than others. The list of sectors is long, and predictions varied, however a few clear winners, such as manufacturing, education, healthcare, agriculture, computer manufacturing and transport seem to emerge (Forbes, 2022; Miao et al., 2021; Ferencz et al, 2022). Based on earlier findings in this review, it can also be hypothesized, that sectors in which AI innovations are the highest are also sectors where AI innovation is needed the most, and these are consequently also the most tradeable sectors.

Focus is therefore once again directed towards global value chains, and the role of international trade in progressing, or hindering AI induced growth. That being said, while hypothesizing around which sectors will be the most affected provides some value, this review is not aware of any papers which aim to quantify differences of magnitude in economic impact between sectors. In Bughin et al. (2018), macroeconomic impact of AI is predicted to be 2.3 times the magnitude for telecom and high-tech sectors than healthcare, which is a substantial difference. It is also noted in the paper, that more firm- and sector-level data is needed for a more comprehensive analysis of sectoral impacts.

### 3.4 Country divide: Lessons of history

Another dimension to consider is the country divide, more specifically the gap between developed and developing countries. Will AI induced growth close the gap or widen it further? Again, lessons of history should be examined, with internet access an obvious choice for a proxy. Again, literature seems to be scarce. According to a 2022 review by the World Bank, some studies have found that internet enabled technologies, such as mobile money in Kenya have managed to increase incomes and reduce poverty. Furthermore, five studies seem to support the theory that internet infrastructure has increased incomes and consumption in Africa. At the same time, it was also found that subsidizing internet access to individuals already possessing a phone contract, but did not use roaming data, had limited effects on well-being (Hjort & Sacchetto, 2022). It is concluded, that while research is scarce, effects seem to be promising.

### 3.5 Country divide: General predictions

Perhaps the most prominent perspective in research relating to sharing of AI induced wealth is that of a country divide, between developed and developing countries. Some findings are favourable towards developing countries. For instance, it is speculated that an AI aided shift towards more skill intensive industries could be utilized by developing countries to transform the structure of their own economies and can penetrate sectors that were not previously available to them, because their resources had been tied up (Kouka & Magallanes, 2022). It is also theorized, that developing countries, which are more integrated into global value chains than developed ones (both backwards and forwards linkages of non-OECD countries is stronger than OECD economies'), will benefit greatly from reductions in international trade frictions (for instance through AI powered translation). In addition, reduced compliance costs and an overall decreased regulatory burden brought forward by AI technology can increase accessibility of trade financing (Jayathilaka, 2022), which in combination with reduced trade barriers could disproportionately benefit especially the least developed countries.

The being said, an overwhelming share of the research in the field points towards a bleak future for developing countries, if we follow the trajectory currently lined out. A common theme that can be identified is control over technology. Those who control the production (for example chip production), and those who control the development of AI technologies, and technologies adjacent to AI, will be the big winners of the next few decades. Firstly, as AI is introduced into economies and societies, those in control of the technology will have a very high economic and political bargaining power against those who don't (Brynjolfsson et al, 2018; Kouka & Magallanes, 2022).

Risks of onshoring also await developing countries. As robotization reduces the need for cheap labour, multinationals could choose to bring production closer to home, which can lead to further decrease on growth potential and comparative advantage for countries which depend highly on FDI through offshoring (Artuc et al, 2023; Spence, 2022). As a result, large labour market adjustments would be needed, at a pace that many countries may not be able to keep up with. Further disparities may be created in developing countries through the widening gap between more advanced, internationally active firms that account for a large portion of exports and small informal firms that account for a large share of low skilled and manual employment (Artuc et al, 2023). Furthermore, as access to up to date, advanced databases becomes essential for competitiveness, SME-s, especially in less advanced economies will suffer (Jayathilaka, 2022). Finally, obstacles of international law also prove to be detrimental. Tariff rates are the highest in developing countries, many of which do not participate in the WTO Information Technology Agreement. This in itself will act as a barrier for adoption of AI technologies via trade, as well as for the development of in-house AI technologies. Bughin et al. (2018) come to similar conclusions, where net GDP impact of AI adoption hovers around 20 percent for advanced economies, but doesn't even reach 10, for the least advanced economies by the year of 2030.

**3.6 Country divide: International trade theory perspective**

A last perspective on wealth sharing can be taken through the only attempt (thus far) at integrating AI developments into international trade theory. Hazari et al (2022) find that AI could reverse trade patterns and lead to factor intensity reversals, and the creation of a Leontief paradox (Hazari et. al, 2022). The finding resonates with the conclusions in Kiyota and Kurokawa (2021), who conclude that factor reversals already exist, based on an analysis of Japanese prefectural data. A factor intensity reversal essentially entails a scenario where the relative abundance of specific factors in production changes in a direction opposite to a country's competitive advantage (Hillman, A.L. and Hirsch, 1979). A historical example of this is the decrease in manufacturing in developed countries and the increase in manufacturing in developing countries at the same time, to a point where we currently have a relative abundance of skilled labour in developing countries, giving them a comparative advantage in manufacturing. A Leontief paradox occurs when a country with a higher capital per worker ratio has a lower capital per labour ratio in exports rather than imports. Put differently, if a country is labour abundant, it would have to have capital intensive exports and imports of labour-intensive goods. The exact implications of a scenario, where developing countries experience factor intensity reversals are hard to predict. That being said generally speaking, a factor reversal for developing countries would imply a shift away from their competitive advantage, namely exports of labour-intensive goods, and also be in line with predictions of increased onshoring by developed countries putting developing countries under high pressure to adapt under a short amount a time.

# 4. Discussion

Perhaps the most overarching conclusion that can be drawn from this review is the need for more research. When it comes to integration of AI into economic theory, we are very much at the beginning stages, and that applies to international trade especially. As for impacts of AI, specific use cases have been discussed in detail, however overarching perspectives weighing benefits against costs and timeframes have been provided to a much lesser extent, rather unsurprisingly so.

Still, some conclusions can be drawn, which are presented below. The path towards transformative AI looks complex, as expected. It seems like certain sectors will both be impacted and drive AI innovation to a larger extent than others, such as manufacturing and high technology. It is also clear, that most of these sectors are highly tradeable, which implies the importance of international trade in driving AI innovation forward. Past disruptions such as the introduction of internet point towards a more gradual adoption of AI, and thus more gradually appearing impacts. Yet, some examples, such as the adoption rate of Bing AI could imply that something entirely different is about to unfold. If certain feedback loops become reality, such as the self-replication of machines, or AI based rapid knowledge enhancement, a scenario of rapid integration of AI into our economies and societies could occur, where human level AI has been developed, and economic growth could reach 30 percent by the year 2100. Several supply chain, regulatory, technological, and knowledge-based bottlenecks stand in way however, some of which could be solved or alleviated by AI technology itself, and some which can't.

Predicting how much wealth AI can create, is difficult, too. We may employ various growth scenarios, starting from the current norm of exponential growth to explosive growth, or even stagnation. Some predict explosive growth to take place by 2100 with a 10 percent probability, while others simulate a global value-added boost of 26 percent by 2030. Taking an international trade perspective, AI technology shows potential in reducing trade frictions, which could in turn reduce supply chain bottlenecks in the way of explosive growth. Still, more overarching studies are needed to be able to state this with any confidence.

The division of wealth created by AI can be viewed on for instance a sectoral level, or with a developed versus developing countries perspective (while recognizing that all countries are different and will need to make decisions on how to best parry the AI revolution based on their economies' each individual characteristics). It can be speculated, that sectors in which AI innovations are the highest, and which contribute the most to AI innovation, will see the most growth thanks to AI technology, and thus will impact national and international value chains in a way where they gain a dominant position. Interestingly, sectors with highest AI innovation are also most integrated into global value chains, and therefore international trade will most likely become a key battling ground where AI will be one of the key factors to ensure competitiveness.

Looking at a historical perspective, we see some proof that introduction of new technologies can create wealth in disadvantaged countries, however more research is needed for conclusive evidence. While AI technology will surely bring some opportunities to such countries, such as the chance towards a shift towards more skill intensive economies, and reduction in trade frictions, the former may also lead to their doom, if they can't adapt on time. Furthermore, those in control of the knowledge, and manufacturing capabilities linked to AI will gain further bargaining power against those who don't. Lastly, as AI replaces increasing shares of the workforce, offshoring may turn to onshoring, further decreasing the competitive advantage for countries exporting labour intensive goods. It is estimated that the net impact advanced economies from AI technologies will be up to 4 times larger than for the least advanced economies.

Integration of AI impacts into international trade theory is still very limited, however a valuable attempt made raises the question: What will happen if the adoption of AI will lead to factor intensity reversals? The implications, again, for developing countries could be more negative than positive.

# 5. Conclusion

What then can we conclude from all of the above? We are currently at a point where research in the field is still much lacking, in other words, we don't yet really know what economic implications AI has in store for us. Some studies point towards possibilities of explosive growth, which would be beneficial for the development of humanity. That being said, it is also clear that a gap in research exists, where AI risk has yet to be integrated into growth models, at attempt, that could bring nuance into the overly positive picture that current frameworks project. Furthermore, based on the current knowledge we have, it seems like the divide between developed and developing countries will increase further as a result of AI adoption.

Given this, the current recommendation is to choose the slow path of AI adoption. Policy makers should ensure that introduction of AI technologies into economies and societies happen in a controlled way. A focus must be put on AI safety, and the several unknowns in that field. Meanwhile, policy frameworks must be developed to make sure that developing countries are better armed for the AI transformation, such as by ensuring their access to AI technologies, and easing trade barriers for ICT goods. Instead of replacing labour fully with AI, we should focus on augmenting it with AI instead, which will also ensure that bargaining power of laborers in less advantaged countries will remain.

# References

Achar, S., 2019. Early Consequences Regarding the Impact of Artificial Intelligence on International Trade. *American Journal of Trade and Policy*, *6*(3), pp.119-126.

Akerman, A., Leuven, E. and Mogstad, M., 2022. Information frictions, internet, and the relationship between distance and trade. *American Economic Journal: Applied Economics*, *14*(1), pp.133-163.

Anderson, J. and Rainie, L., 2018. 3. Improvements ahead: How humans and AI might evolve together in the next decade. *Pew Research Center: Internet, Science & Tech*.

Artuc, E., Bastos, P., Copestake, A. and Rijkers, B., 2022. Robots and trade: Implications for developing countries. In *Robots and AI* (pp. 232-274). Routledge.

Bryant, M. (2011) "20 years ago today, the World Wide Web opened to the public," *The Next Web*, 6 August.

Brynjolfsson, E., Hui, X. and Liu, M., 2019. Does machine translation affect international trade? Evidence from a large digital platform. *Management Science*, *65*(12), pp.5449-5460.

Bughin, J., Seong, J., Manyika, J., Chui, M. and Joshi, R., 2018. Notes from the AI frontier: Modeling the impact of AI on the world economy. *McKinsey Global Institute*, *4*.

Dafoe, A., 2018. AI governance: a research agenda. *Governance of AI Program, Future of Humanity Institute, University of Oxford: Oxford, UK*, *1442*, p.1443.

Davidson, T., 2021. Could advanced AI drive explosive economic growth. *Open Philanthropy*

Ferencz, J., González, J.L. and García, I.O., 2022. Artificial Intelligence and international trade: Some preliminary implications. *OECD publishing*.

Forbes (2022) *Council post: 16 industries and functions that will benefit from AI in 2022 and beyond*, *Forbes*. Forbes Magazine. Available at: https://www.forbes.com/sites/forbestechcouncil/2022/01/13/16-industries-and-functions-that-will-benefit-from-ai-in-2022-and-beyond/?sh=38076131477d.

Gomez-Herrera, E., Martens, B. and Turlea, G., 2014. The drivers and impediments for cross-border e-commerce in the EU. *Information Economics and Policy*, *28*, pp.83-96.

Hazari, B., Lai, J.T. and Mohan, V., 2022. A Note on the Implications of Automation and Artificial Intelligence for International Trade. *Arthaniti: Journal of Economic Theory and Practice*, p.09767479221129186.

Hillman, A.L. and Hirsch, S., 1979. Factor intensity reversals: Conceptual experiments with traded goods aggregates. *Weltwirtschaftliches Archiv*, *115*(2), pp.272-283.

Hjort, J. and Sacchetto, C. (2022) *Can internet access lead to improved economic outcomes?*, *World Bank Blogs*. Available at: https://blogs.worldbank.org/digital-development/can-internet-access-lead-improved-economic-outcomes.

Jayathilaka, U.R., 2022. The Role of Artificial Intelligence In Accelerating International Trade: Evidence From Panel Data Analysis. *Reviews of Contemporary Business Analytics*, *5*(1), pp.1-15.

Jones, E., 2023. Digital disruption: artificial intelligence and international trade policy. *Oxford Review of Economic Policy*, *39*(1), pp.70-84.

Kiyota, K. and Kurokawa, Y., 2022. Factor intensity reversals redux: Feenstra is right!. *Review of International Economics*, *30*(4), pp.885-914.

Kouka, M. and Magallanes, M.C., 2022. Technological Trivergence and International Trade: A Literature Review. *Harvard University*.

Kriegman, S., Blackiston, D., Levin, M. and Bongard, J., 2021. Kinematic self-replication in reconfigurable organisms. *Proceedings of the National Academy of Sciences*, *118*(49), p.e2112672118

Krishnaswamy, S. (no date) *Challenges and strategies to achieve full automation in semiconductor assembly and Test*, *SEMI*. Available at: https://semi.org/en/about/AMAT.

Langenkamp, M. and Yue, D.N., 2022, July. How Open Source Machine Learning Software Shapes AI. In *Proceedings of the 2022 AAAI/ACM Conference on AI, Ethics, and Society* (pp. 385-395).

Manyika, J., Chui, M., Miremadi, M., Bughin, J., George, K., Willmott, P. and Dewhurst, M., 2017. A future that works: AI, automation, employment, and productivity. *McKinsey Global Institute Research, Tech. Rep*, *60*, pp.1-135.

Mateos-Garcia, J. (2021) *The Economics of Artificial Intelligence today*, *The Gradient*. The Gradient. Available at: https://thegradient.pub/the-economics-of-ai-today/

Mauran, C. (2023) "Bing, yes that Bing, now has 100 million daily users," *Mashable*, 9 March. Available at: https://mashable.com/article/microsoft-bing-100-million-daily-users-ai-chatbot.

Miao, F., Holmes, W., Huang, R. and Zhang, H., 2021. *AI and education: A guidance for policymakers*. UNESCO Publishing.

Spence, M., 2022. Automation, Augmentation, Value Creation & the Distribution of Income & Wealth. *Daedalus*, *151*(2), pp.244-255.

Stackpole, B. (2020) 5 supply chain technologies that deliver competitive advantage, MIT Sloan. Available at:https://mitsloan.mit.edu/ideas-made-to-matter/5-supply-chain-technologies-deliver-competitive-advantage.

The New York Times (1999) "Internet Users Now Exceed 100 Million," *The New York Times*, 12 November, 1999.

Trammell, P. and Korinek, A., 2023. Economic growth under transformative AI: A guide to the vast range of possibilities for output growth, wages, and the labor share. *Global Priorities Institute*.